# Incommensurate spin-density wave order in electron-doped BaFe$_2$As$_2$ superconductors

D. K. Pratt, M. G. Kim, A. Kreyssig, Y. B. Lee, G. S. Tucker, A. Thaler, W. Tian,

J. L. Zarestky, S. L. Bud'ko, P. C. Canfield, B. N. Harmon,

A. I. Goldman, R. J. McQueeney

*Ames Laboratory, U.S. DOE and Department of Physics and Astronomy, Iowa State University, Ames, Iowa 50011, USA*

**Abstract.** Neutron diffraction studies of Ba(Fe$_{1-x}$Co$_x$)$_2$As$_2$ reveal that commensurate antiferromagnetic order gives way to incommensurate magnetic order for Co compositions between $0.056 < x < 0.06$. The incommensurability has the form of a small transverse splitting $(0, \pm\varepsilon, 0)$ from the nominal commensurate antiferromagnetic propagation vector $\mathbf{Q}_{AFM} = (1, 0, 1)$ (in orthorhombic notation) where $\varepsilon \approx 0.02 - 0.03$ and is composition dependent. The results are consistent with the formation of a spin-density wave driven by Fermi surface nesting of electron and hole pockets and confirm the itinerant nature of magnetism in the iron arsenide superconductors.

Unconventional superconductivity is often associated with the pairing of electrons via spin fluctuations that appear close to a magnetic ordering instability. In this respect, the nature and origin of the magnetic instability itself is an important ingredient of any theory of superconductivity. In the iron arsenide compounds, the magnetism has been discussed from two limits; an itinerant and a local moment limit. The parent $AE$Fe$_2$As$_2$-based superconductors ($AE$ = Ca, Sr, Ba) are antiferromagnetic (AFM) metals, which suggests that an itinerant description is an appropriate starting point. AFM order is observed with a commensurate magnetic propagation vector $\mathbf{Q}_{AFM} = (1, 0, 1)$ (expressed in orthorhombic notation) in a variety of iron arsenide compounds by neutron and x-ray resonant magnetic diffraction.[1-9] The small ordered moments measured in these systems ($< 1$ $\mu_B$) also favor an itinerant description. In principle, the propagation vector of the AFM order itself, $\mathbf{Q}_{AFM}$, should further strengthen the case for itinerant magnetism, as both band structure calculations [10,11] and angle-resolved photoemission data [12-14] display Fermi surface nesting between electron and hole pockets with a nesting vector close to $\mathbf{Q}_{AFM}$. Here we define an itinerant spin-density wave (SDW) as magnetic order resulting from an instability due to Fermi surface nesting, with the best known example being the incommensurate (IC) SDW order observed in Cr metal.[15] However, the commensurate (C) AFM order observed at $\mathbf{Q}_{AFM}$ can also be described within a local moment picture that may become relevant in the presence of moderately large electronic correlations and can be quantified, for example, in terms of the $J_1$ - $J_2$ Heisenberg model where $J_2 > 2J_1$.[16]

Detailed band structure calculations of the magnetic susceptibility in the iron arsenides predict that the Fermi surface nesting condition can result in either C-SDW order at $\mathbf{Q}_{AFM}$, or IC-SDW order with a propagation vector $\boldsymbol{\tau} = \mathbf{Q}_{AFM} + \boldsymbol{\varepsilon}$ where ε is a small incommesurability.[17,18] Although the observation of IC magnetic order with a propagation vector similar to that predicted by band structure calculations would clearly favor an itinerant SDW description of the $AE$Fe$_2$As$_2$ system, detailed magnetic diffraction studies have observed only C-AFM order with a propagation vector $\mathbf{Q}_{AFM}$ in several $AE$Fe$_2$As$_2$ systems including the parent compounds [1-3] and doped compounds: Ba(Fe$_{1-x}$Co$_x$)$_2$As$_2$,[4-6] Ba(Fe$_{1-x}$Ni$_x$)$_2$As$_2$,[7] Ba(Fe$_{1-x}$Rh$_x$)$_2$As$_2$,[8] Ba(Fe$_{1-x}$Ru$_x$)$_2$As$_2$,[9] Ba$_{1-x}$K$_x$Fe$_2$As$_2$.[19] Incommensurability has been claimed in Ba(Fe$_{1-x}$Co$_x$)$_2$As$_2$ by local probes such as $^{75}$As nuclear magnetic resonance [20], $^{57}$Fe Mössbauer [21], and muon spin resonance [22] measurements. However, detailed measurements of the AFM ordering by both neutron and high resolution x-ray resonant magnetic diffraction have found no incommensurability in Ba(Fe$_{1-x}$Co$_x$)$_2$As$_2$ up to $x = 0.047$.[6]

In this Letter, neutron diffraction data demonstrate that IC magnetic order does indeed develop near optimally doped compositions of Ba(Fe$_{1-x}$Co$_x$)$_2$As$_2$ with $x \geq 0.056$, just before long-range magnetic ordering is completely suppressed at $x \approx 0.06$. The IC propagation vector $\boldsymbol{\tau} = \mathbf{Q}_{AFM} + (0, \varepsilon, 0)$ corresponds to a transverse splitting ($\varepsilon \approx 0.02 - 0.03$) that depends on composition. The direction and magnitude of the observed IC splitting is consistent with calculations of the generalized susceptibility determined by density functional theory, allowing us to conclude that static magnetism and the spin fluctuations for superconducting compositions are tied to an itinerant SDW instability.

Neutron diffraction measurements were performed at Oak Ridge National Laboratory using the High Flux Isotope Reactor's HB1A spectrometer. Single-crystals of Ba(Fe$_{1-x}$Co$_x$)$_2$As$_2$ with the following compositions (and masses); $x$ = 0.054 (124 mg), 0.056 (248 mg), 0.057 (73 mg), 0.059 (136 mg), and 0.062 (106 mg). The sample compositions were determined through a series of characterization measurements including resistivity, magnetization, and wavelength dispersive spectroscopy.[23] All samples are orthorhombic below the tetragonal-orthorhombic transition temperature ($T_S$) and the data are discussed in terms of the orthorhombic indexing $\mathbf{Q} = \left(\frac{2\pi}{a}H, \frac{2\pi}{b}K, \frac{2\pi}{c}L\right)$ where $a \approx b \approx 5.6$ Å and $c \approx 13$ Å. The experimental configuration for the triple-axis spectrometer included the horizontal collimation 48' - 48' - 40' - 68' with $E_i$ = 14.7 meV and two graphite filters to eliminate higher harmonic contamination from the incident beam. All samples have resolution-limited mosaic full-widths of < 0.4 degrees. Measurements were performed in the vicinity of $\mathbf{Q}_{AFM}$ = (1, 0, 1) and (1, 0, 3) with samples mounted in a closed-cycle refrigerator for low temperature studies. In order to search for incommensurability in the direction of the two in-plane orthorhombic axes, samples were studied in two scattering planes; in the ($H$, 0, $L$) plane, allowing the search for IC splitting along the orthorhombic **a** axis ([$H$, 0, 0] is referred to as the longitudinal direction), and in the ($\xi$, $K$, 3$\xi$) or ($\xi$, $K$, $\xi$) planes, allowing the search for incommensurability along the **b** axis ([0, $K$, 0], transverse direction). The geometry of reciprocal space scans is shown in Fig. 1(a).

Typical transverse [0, $K$, 0] and longitudinal [$H$, 0, 0] neutron diffraction scans are shown in Figs. 1(b) and (c) for the $x = 0.059$ sample at the superconducting transition temperature ($T_c$), where $T_c$ is below the AFM ordering temperature ($T_N$). The observation of a pair of Bragg peaks located symmetrically at positions (0, $\pm\varepsilon$, 0) around $\mathbf{Q}_{AFM}$ in the transverse scan clearly indicates IC magnetic order for this composition. The longitudinal scan reveals a single magnetic peak due to residual intensity from the overlap of the transversely split Bragg peaks. Most importantly, no additional longitudinal splitting is observed, therefore IC magnetic order is present with propagation vector $\boldsymbol{\tau} = (1, \varepsilon, 1) = \mathbf{Q}_{AFM} + (0, \varepsilon, 0)$, as illustrated in Fig. 1(a).

We now turn to *ab initio* calculations of the magnetic susceptibility in order to show that the observed IC-AFM order can be understood as a SDW driven by Fermi surface nesting. Previous electronic structure calculations using density functional theory show maxima in the generalized spin susceptibility away from $\mathbf{Q}_{AFM}$ in doped $AE$Fe$_2$As$_2$ compounds and therefore point to a tendency for IC-SDW order.[17,18] To gain insight into potential incommensurability at doping levels where we observe static IC-AFM order, we performed calculations of the generalized bare susceptibility employing the full-potential linearized augmented plane wave (FPLAPW) method,[24] with a local density functional.[25] We used $R_{MT}K_{max} = 8.0$ and $R_{MT} = 2.4$, 2.2 and 2.2 for Ba, Fe and As respectively. To obtain self-consistency we chose 550 **k**-points in the irreducible Brillouin zone and used 0.01 mRy/cell as the total energy convergence criteria. The virtual crystal approximation was used to consider Co doping effects and the whole reciprocal unit cell is divided into 80*80*80 parallelepipeds, corresponding to 34061 irreducible **k**-points. The results of our calculations of the generalized susceptibility for electron-doping with $x = 0.05$ show splitting in the transverse direction and a single peak in the longitudinal direction in Figs. 1(d) and 1(e), respectively, consistent with other doping dependent calculations mentioned above.[17,18] The *ab initio* calculations therefore show a tendency for IC-SDW order with propagation vector

$\tau \approx \mathbf{Q}_{AFM} + (0, \varepsilon, 0)$ in agreement with experimental observations.

Figure 2 shows the transverse [0, $K$, 0] scans through (1, 0, 3) for other compositions and temperatures $x = 0.054, 0.056, 0.057$, and $0.059$ with $T < T_c$, $T \approx T_c$, $T_c < T < T_N$, and $T > T_S$. The scans performed at temperatures above $T_S$ serve as an estimate of the background. Details of the magnetic structure such as the propagation vector, peak widths, and integrated intensities were determined by Gaussian fits to the scans shown in Fig. 2. The transverse [0, K, 0] scans show only a single resolution-limited peak for the x = 0.054 sample and, combined with x-ray resonant magnetic diffraction results from Kim *et al.* [6], establish stripe-type C-AFM order at $Q_{AFM}$ for all Co compositions below approximately 0.054. Broad peaks split in the transverse direction are observed in [0,$K$,0] scans for $x = 0.056, 0.057$, and $0.059$, clearly establishing the transition to an IC magnetic phase with propagation vector $\tau$. For the $x = 0.056$ sample, both C and IC peaks are observed, suggesting that the transition is first-order in its dependence on Co concentration and the $x = 0.056$ composition is close to the phase boundary. Figure 2(c) shows that the lineshapes at (1, 0, 3) and (1, 0, 1) positions are equivalent with an integrated intensity ratio of 0.36(9), close to that expected for collinear and C-AFM order with the magnetic moment pointing along the **a** axis. The IC-AFM structure is therefore most likely also collinear, and not helical or cycloidal, with magnetic moments along the **a** axis. No signatures of higher harmonics have been observed, signifying a sinusoidal modulation of the moment size along the **b** direction.

We now discuss the temperature dependence of transverse scans shown in Fig. 2. For $x = 0.054$ [Fig. 2(a)], the suppression of the integrated intensity (magnetic order parameter) below $T_c$ indicates the competition of C-AFM with superconductivity, as reported previously.[5,26,27] The magnetic intensity in Figs. 2(b) - (d) show a similar suppression below $T_c$ implying that the IC state also competes with superconductivity.

Figure 3(a) shows the experimental phase diagram of $Ba(Fe_{1-x}Co_x)_2As_2$ delineating regions of magnetic order, superconductivity, and structural phases as based on previous studies. [9-11,23-25]. This work, summarized in Figs. 3(b) - (d), has allowed us to outline regions of C and IC magnetic order in the phase diagram. Fig. 3(b) shows the evolution from C (at $x = 0.054$) to IC order (from $x = 0.056$ - $0.059$) in transverse scans performed at $T \approx T_c$. The $x = 0.062$ sample has no detectable magnetic order. The composition dependence of both the integrated magnetic intensity and incommensurability is plotted in Fig. 3(c) at $T \approx T_c$, again highlighting that the first-order transition to IC magnetic order occurs at $x \approx 0.056$ in the limit where the magnetic intensity (moment size) is very small. The incommensurability grows slightly at the higher compositions, reaching a value of 0.030(2) at $x = 0.059$. Figure 3(d) displays the temperature dependence of the integrated intensity of IC magnetic Bragg peaks for the $x = 0.056$ sample, which has the characteristic suppression in the superconducting state, as alluded to above. Figure 3(d) also plots the incommensurability parameter, $\varepsilon$, of the $x = 0.056$ sample as a function of temperature. The incommensurability is finite just below $T_N$ [$\varepsilon = 0.018(5)$] and grows until $T_c$ is reached. Below $T_c$, the value of $\varepsilon = 0.025(3)$ is relatively constant suggesting that the coupling of superconductivity and magnetism may act to pin the incommensurability parameter.

The magnetic phase diagram shown in Fig. 3(a) contains a first-order C-to-IC transition with electron-doping in Ba(Fe$_{1-x}$Co$_x$)$_2$As$_2$ that bears a strong similarity to the alloys of the canonical SDW system, Cr. Pure Cr orders into an IC-SDW state that is driven by nesting between electron and hole Fermi surfaces whose areas are slightly mismatched.[15] Electron-doping of Cr (in this case by alloying with Mn [28] or Ru [29]) equalizes the Fermi surface areas and results in a first-order transition to C-SDW order. This simple picture considers only the Fermi surface topology and the free energy of competing C and IC-SDW states and has led to a detailed theoretical understanding of the magnetic phase diagram of Cr alloys.[30]

The development of C or IC-SDW order has also been studied in the iron arsenides using an effective two-dimensional, two-band Ginzburg-Landau approach.[10,31] In a spirit similar to Cr, IC-SDW order is favored when nesting occurs between electron and hole pockets having circular cross-sections of unequal area at the Fermi level. The introduction of more realistic elliptical electron pockets favor C-SDW order as long as the electron and hole pocket areas are not too strongly mismatched, as is the case for the parent compounds. However, even with elliptical electron pockets, doping detunes the two pockets and eventually results in a mismatch that favors IC-SDW order. This analysis suggests that Fermi surface nesting is a crucial factor in stabilizing both C and IC phases in the magnetic phase diagram of the $AE$Fe$_2$As$_2$ compounds.

Unlike Cr, the doped iron arsenides are superconductors, and both C and IC-SDW order are observed to coexist with superconductivity. Ginzburg-Landau models [10,31] indicate that the competition and coexistence of superconductivity with either C or IC-SDW order is much more likely with an unconventional $s^{+-}$ pairing symmetry. Thus, a simple two-band approach appears to capture many of the essential features of the phase diagram of the $AE$Fe$_2$As$_2$ arsenides in terms of Fermi surface nesting, C and IC-SDW order, and unconventional $s^{+-}$ superconductivity. The resulting theoretical phase diagram [31] bears close resemblance to the experimental diagram in Fig. 3(a).

In summary, we have observed a first-order transition from commensurate to incommensurate antiferromagnetic order with electron-doping in Ba(Fe$_{1-x}$Co$_x$)$_2$As$_2$. The combination of this experimental finding, density functional theory calculations of the generalized susceptibility, and model calculations of Vorontsov et al.,[31] establish a very strong case for itinerant SDW order and unconventional superconductivity in the $AE$Fe$_2$As$_2$ arsenides.

ACKNOWLEDGMENTS. We acknowledge valuable discussions with J. Schmalian, R. M. Fernandes, T. Brueckel, and R. Hermann. This work was supported by the Division of

REFERENCES

[1]  A. I. Goldman *et al.*, Phys. Rev. B **78**, 100506 (2008).
[2]  J. Zhao *et al.*, Physical Review B **78**, 140504 (2008).
[3]  Q. Huang *et al.*, Phys. Rev. Lett. **101**, 257003 (2008).
[4]  C. Lester *et al.*, Phys. Rev. B **79**, 144523 (2009).
[5]  R. M. Fernandes *et al.*, Physical Review B **81**, 140501 (2010).
[6]  M. G. Kim *et al.*, Phys. Rev. B **82**, 180412 (2010).
[7]  M. Wang *et al.*, Phys. Rev. B **81**, 174524 (2010).
[8]  A. Kreyssig *et al.*, Phys. Rev. B **81**, 134512 (2010).
[9]  M. G. Kim *et al.*, Phys. Rev. B **83**, 054514 (2011).
[10] A. B. Vorontsov, M. G. Vavilov, and A. V. Chubukov, Phys. Rev. B **79**, 060508 (2009).
[11] V. Cvetkovic and Z. Tesanovic, Phys. Rev. B **80**, 024512 (2009).
[12] J. Fink *et al.*, Phys. Rev. B **79**, 155118 (2009).
[13] P. Vilmercati *et al.*, Phys. Rev. B **79**, 220503 (2009).
[14] C. Liu *et al.*, Nat. Phys. **6**, 419 (2010).
[15] E. Fawcett *et al.*, Rev. Mod. Phys. **66**, 25 (1994).
[16] Q. Si and E. Abrahams, Phys. Rev. Lett. **101**, 076401 (2008).
[17] J. T. Park *et al.*, Phys. Rev. B **82**, 134503 (2010).
[18] S. Graser *et al.*, Phys. Rev. B **81**, 214503 (2010).
[19] H. Chen *et al.*, Europhys. Lett. **85**, 17006 (2009).
[20] Y. Laplace *et al.*, Phys. Rev. B **80**, 140501 (2009).
[21] P. Bonville and et al., Europhys. Lett. **89**, 67008 (2010).
[22] P. Marsik *et al.*, Phys. Rev. Lett. **105**, 057001 (2010).
[23] N. Ni *et al.*, Phys. Rev. B **78**, 214515 (2008).
[24] P. Blaha *et al.*, *WIEN2k, An Augmented Plane Wave + Local Orbitals Program for Calculation Crystal Properties* (TU Wien, Austria, 2001).
[25] J. P. Perdew and Y. Wang, Phys. Rev. B **45**, 13244 (1992).
[26] D. K. Pratt *et al.*, Phys. Rev. Lett. **103**, 087001 (2009).
[27] A. D. Christianson *et al.*, Phys. Rev. Lett. **103**, 087002 (2009).
[28] B. J. Sternlieb *et al.*, Phys. Rev. B **50**, 16438 (1994).
[29] R. S. Eccleston and et al., J. Phys.: Condens. Matter **8**, 7837 (1996).
[30] R. S. Fishman and S. H. Liu, Phys. Rev. B **48**, 3820 (1993).
[31] A. B. Vorontsov, M. G. Vavilov, and A. V. Chubukov, Phys. Rev. B **81**, 174538 (2010).

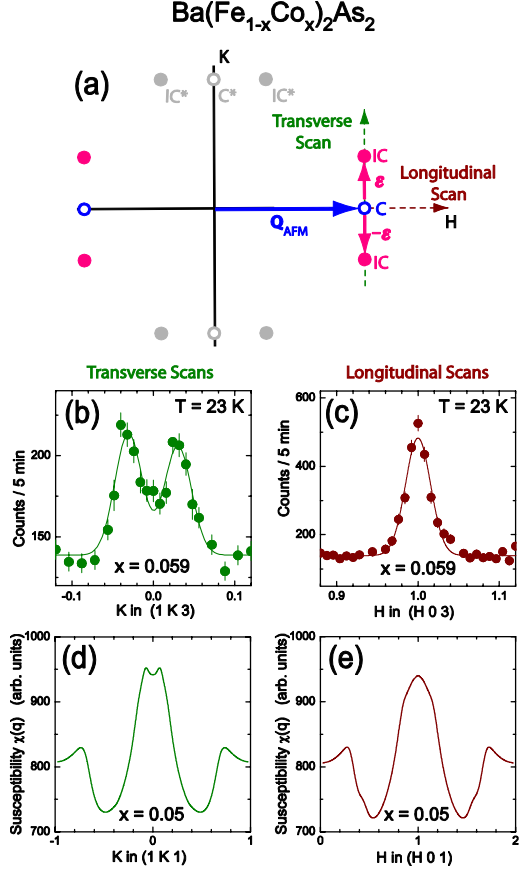

FIG. 1. (a) Reciprocal space plane with $L = odd$ indicating commensurate (C, empty circles) and incommensurate (IC, filled circles) magnetic Bragg peak positions at $\mathbf{Q}_{AFM} = (1, 0, L = odd)$ and $\boldsymbol{\tau} = (1, \pm\varepsilon, L = odd)$, respectively, in orthorhombic notation. The size of the incommensurability parameter $\varepsilon$ is exaggerated for clarity. Shaded points labeled C* and IC* show the location of magnetic Bragg peaks at both $(0, 1, L = odd)$ and $(\pm\varepsilon, 1, L = odd)$, respectively, that are present due to orthorhombic twinning. Dashed arrows illustrate the direction of longitudinal and transverse neutron diffraction scans along the $[H, 0, 0]$ and $[0, K, 0]$ directions, respectively. Raw (b) transverse and (c) longitudinal neutron diffraction scans for $Ba(Fe_{0.941}Co_{0.059})_2As_2$ at $T = 23$ K $\approx T_c$. The lines are Gaussian fits to the data as described in the text. *Ab initio* calculations of the generalized susceptibility in the (d) transverse and (e) longitudinal directions thru $\mathbf{Q}_{AFM}$.

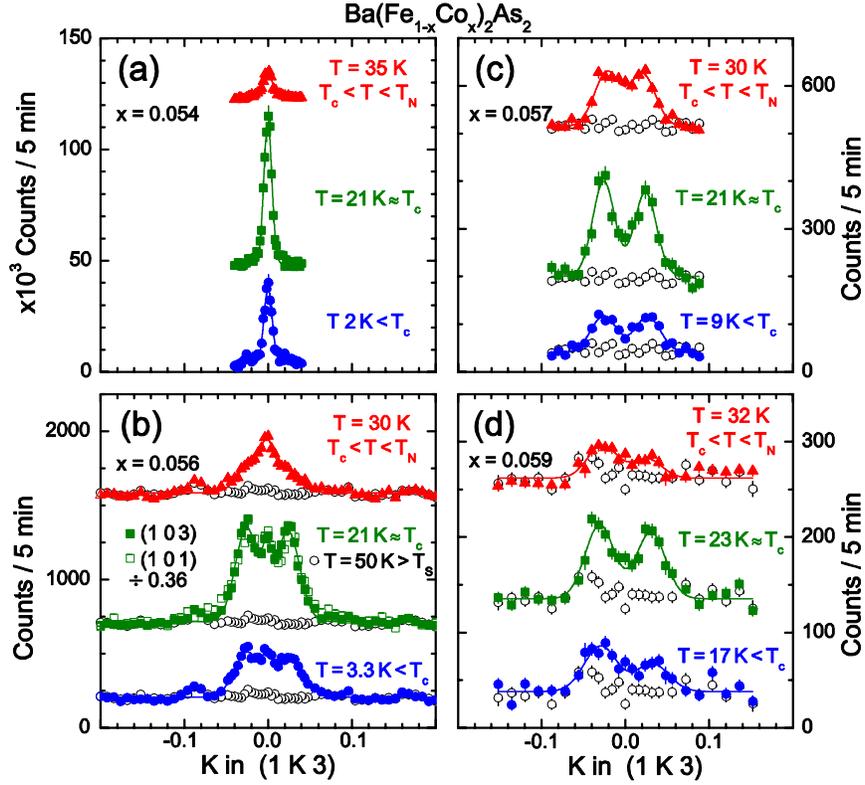

FIG. 2. Transverse neutron diffraction scans at temperatures $T < T_c$ (blue circles), $T \approx T_c$ (green squares), $T_c < T < T_N$ (red triangles), and $T > T_S$ (empty circles) for Ba(Fe$_{1-x}$Co$_x$)$_2$As$_2$ with $x =$ (a) 0.054, (b) 0.056, (c) 0.057, and (d) 0.059. Scans with $T > T_S$ are an estimate of the background. All scans are performed through the (1, 0, 3) position except the empty squares in (c), which are measured through (1, 0, 1) and with the intensity divided by a factor of 0.36. The lines are Gaussian fits to the data.

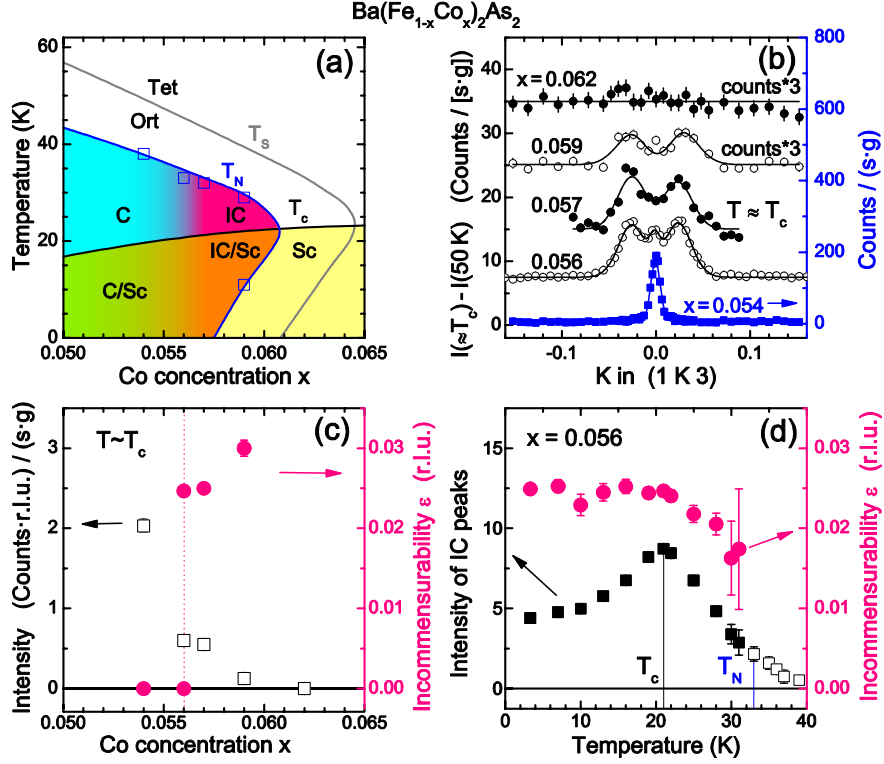

FIG. 3. (a) Experimental phase diagram for Ba(Fe$_{1-x}$Co$_x$)$_2$As$_2$ showing commensurate (C) and incommensurate (IC) antiferromagnetic order below $T_N$. Tetragonal (Tet) and orthorhombic (Ort) phases are separated by the phase line at $T_S$. Finally, superconductivity (Sc) appears below $T_c$ and can coexist with both commensurate (C/Sc) and incommensurate (IC/Sc) magnetic order. Open squares represent the magnetic phase transition temperatures determined in this study. (b) Background subtracted transverse neutron diffraction scans for $x = 0.054$, 0.056, 0.057, 0.059, and 0.062 at $T \approx T_c$. Scans are offset vertically and scaled (where noted) for clarity. (c) Integrated intensity (squares) and incommensurability parameter $\varepsilon$ (circles) as a function of Co concentration at $T \approx T_c$. (d) Integrated intensity and incommensurability parameter $\varepsilon$ as a function of temperature for $x = 0.056$. Open squares represent the total intensity for the broad magnetic Bragg peak above $T_N$. In (c) and (d), bars represent standard deviations from fits.